# Layer-dependent evolution of electronic structures and correlations in rhombohedral multilayer graphene

Yang Zhang[1,2,†], Yue-Ying Zhou[1,2,†], Shihao Zhang[1,†], Hao Cai[1,2,†], Ling-Hui Tong[1,2], Wei-Yu Liao[1,2], Ruo-Jue Zou[1,2], Si-Min Xue[1,2], Yuan Tian[1], Tongtong Chen[3], Qiwei Tian[1], Chen Zhang[1], Yiliu Wang[1], Xuming Zou[1], Xingqiang Liu[4], Yuanyuan Hu[4], Ya-Ning Ren[5], Li Zhang[1], Lijie Zhang[1], Wen-Xiao Wang[3,*], Lin He[5], Lei Liao[4], Zhihui Qin[1,*], and Long-Jing Yin[1,2,*]

[1] *Key Laboratory for Micro/Nano Optoelectronic Devices of Ministry of Education & Hunan Provincial Key Laboratory of Low-Dimensional Structural Physics and Devices, School of Physics and Electronics, Hunan University, Changsha 410082, China*

[2] *Research Institute of Hunan University in Chongqing, Chongqing 401120, China*

[3] *College of Physics and Hebei Advanced Thin Films Laboratory, Hebei Normal University, Shijiazhuang, Hebei 050024, China*

[4] *College of Semiconductors (College of Integrated Circuits), Hunan University, Changsha, 410082, China*

[5] *Center for Advanced Quantum Studies, Department of Physics, Beijing Normal University, Beijing 100875, China*

[†]These authors contributed equally to this work.

*Corresponding authors: wangwx@hebtu.edu.cn; zhqin@hnu.edu.cn; yinlj@hnu.edu.cn

**The recent discovery of superconductivity and magnetism in trilayer rhombohedral graphene (RG) establishes an ideal, untwisted platform to study strong correlation electronic phenomena. However, the correlated effects in multilayer RG have received limited attention, and, particularly, the evolution of the correlations with increasing layer number remains an unresolved question. Here, we show the observation of layer-dependent electronic structures and correlations—under surprising liquid nitrogen temperature—in RG multilayers from 3 to 9 layers by using scanning tunneling microscopy and spectroscopy. We explicitly determine layer-enhanced low-energy flat bands and interlayer coupling strengths. The former directly demonstrates the further flattening of low-energy bands in thicker RG, and the latter indicates the presence of varying interlayer interactions in RG multilayers. Moreover, we find significant splittings of the flat**

**bands, ranging from ~50–80 meV, at 77 K when they are partially filled, indicating the emergence of interaction-induced strongly correlated states. Particularly, the strength of the correlated states is notably enhanced in thicker RG and reaches its maximum in the six-layer, validating directly theoretical predictions and establishing abundant new candidates for strongly correlated systems. Our results provide valuable insights into the layer dependence of the electronic properties in RG and demonstrate it as a suitable system for investigating robust and highly accessible correlated phases.**

The low-energy flat band constructed in two-dimensional (2D) van der Waals materials has emerged as a new playground for exploring and engineering strongly correlated physics. The most representative family of 2D flat-band systems is twisted moiré heterostructures[1-4], in which moiré superlattices can lead to the emergence of extremely flat minibands that host strong electron–electron interactions[5]. A variety of exotic quantum phenomena driven by flat bands have been consecutively demonstrated in different types of moiré materials, such as twisted bilayer graphene[6-8], twisted multilayer graphene[9-13], trilayer graphene/hexagonal boron nitride (hBN) heterostructures[14-17], and twisted transition metal dichalcogenides[18,19]. Although extensive studies have been carried out in this rising field, understanding the fundamental physics of the flat-band-induced correlated phases is still precluded. In light of this, expanding the family of 2D flat-band systems and studying the correlated states in even simpler materials are highly desired. Rhombohedral graphene (RG; or *ABC*-stacked graphene) is one such promising flat-band system[20-24], as it naturally possesses nearly flat bands at the charge neutrality point (CNP), which arise from the unique energy dispersion of $E \sim k^N$ (where $N$ is the layer number)[25]. Unlike moiré heterostructures, RG does not require twisting and thus exhibits higher accessibility and improved structural uniformity. More remarkably, the flat-band electronic correlations in RG are predicted to exhibit a significant dependence on the number of graphene layers[26,27]. Recently, flat-band-induced superconductivity and magnetism in RG trilayer[28,29] and correlated states in RG thick-layers[30-34] have been discovered, indicating the suitability of RG system for exploring flat-band correlation physics. However, the flat-band physics in RG multilayers have yet to be sufficiently examined experimentally, and their layer dependence of the electronic structures and correlated effects remains unexplored.

Here we fabricated high-quality RG of 3 to 9 layers and report the observation of layer-dependent electronic structures and correlated states in these systems under surprising liquid nitrogen temperature through scanning tunneling microscopy and spectroscopy (STM and STS) measurements. We observe sharp density of states (DOS) peaks induced by flat bands at the CNP, as well as DOS peaks arising from remote bands in tunneling spectra for all studied $N < 10$ RG. These typical characteristics of RG enable us to accurately examine the layer dependence of the electronic structures. We give direct experimental evidence for layer number-induced extension of flat bands in RG and extract layer-enhanced nearest-neighbor interlayer coupling strength $\gamma_1$, a key parameter that determines the band structures. Furthermore, via measuring filling-varied STS, we observe large splittings of the flat-band peaks at partial fillings with the splitting energies ranging from ~50–80 meV at 77 K, indicating the existence of robust interaction-driven correlated states in RG multilayers. In particular, the strength of the observed correlated states exhibits a layer-dependent enhancement and shows a maximum in the 6-layer (6L) RG, directly confirming the theory predictions. We also perform theoretical calculations to interpretate our findings.

Our multilayer RG samples were stacked on hBN layers on SiO$_2$/Si substrate using van der Waals stacking technique (Methods). The STM/STS measurements were performed at 77 K. Figure 1a shows the schematic of the device scheme in STM measurement and the stacking configuration of RG. For RG multilayers, each layer is *AB*-stacked with the nearest-neighbor and next nearest-neighbor layers, forming an *ABC* stacking order. We fabricated multiple multilayer RG devices including 3L-7L and 9L (see Supplementary Figs. 1 and 2). Figure 1b,c show the large-area STM topographic image and atomic-resolution image of the 6L RG sample, which display relatively flat and impurity-free features. Similar surface topography is observed in all RG samples, indicating the high-quality of our fabricated RG. Besides, there are no large moiré superlattices observed in our RG multilayers, indicating the nonalignment between the RG and underlying hBN (we intentionally misaligned them during fabrications). The high sample quality and the moiréless feature of the fabricated RG provide an unprecedented platform to investigate the intrinsic electronic properties of RG multilayers.

Figure 1d shows a representative STS curve (i.e., *dI/dV-V* spectrum reflecting the local DOS of electrons at the measuring point) for the 6L RG. A sharp and pronounced DOS peak (marked by FB) is observed near the Fermi level in the spectrum. This DOS characteristic peak originates from the nearly flat band of RG localized on the top layer, which is similar to what has been extensively observed in trilayer RG[35-39]. The flat-band peak and the entire tunneling spectrum exhibit a high degree of spatial homogeneity over hundreds of nanometers on the sample surface, with only a small variation of ~±3 meV in the position of the flat band (see Fig. 1e). It is noteworthy that appreciable local variations in doping can still be observed at the micrometer scale, resulting in different fillings of the flat band, which presents an opportunity for investigating its correlation effects (see below). Besides the flat-band peak, there are several minor DOS peaks (marked by ±RB) at higher energies in the tunneling spectrum of Fig. 1d, which are caused by van Hove singularities located at the remote band edges of RG multilayers. Below we will show that these remote-band peaks can be utilized to determine fundamental parameters of RG band structures.

The high-quality samples with varying layer numbers enable us to access the layer-dependent electronic properties of RG multilayers. Figure 2a shows *dI/dV* spectra measured in the RG of different layer numbers. The flat-band peak and remote-band peaks are observed in all studied RG multilayers, which are well replicated by the calculated LDOS (Fig. 2d). We first focus on the flat-band state. The width of the flat-band state ranges from 43–53 meV in our RG (upper panel of Fig. 2b) by measuring the full width at half maximum (FWHM) of the DOS peak. These values of the flat-band width are comparable to, and even smaller than those obtained on HOPG—a typical high-quality substrate—under identical experimental conditions (see Supplementary Fig. 3). This once again demonstrates the exceptional quality of our RG samples. A notable characteristic for the flat-band state is that its DOS peak is stronger in thicker RG. This feature can be quantitatively evaluated by comparing the relative intensities of the DOS peaks between the flat-band state and the first remote-band states in different layers. As depicted in Fig. 2b (lower panel), the measured relative intensity, $I_{FB}/I_{RB_1}$, increases with the RG layers. This behavior can be explained by the further flattening of the bands at the CNP in thicker RG. For RG multilayers, the energy bands at the CNP exhibit a local flatness near the *K*-point of the Brillouin zone. As the number of layers increases, this flat-band region expands in momentum space, resulting in larger DOS at the CNP for thicker RG (Fig. 2d,e)—which may favor stronger interactions. Although this phenomenon has been well predicted by theoretical models[21,27], there is currently no direct experimental confirmation. Our layer-varied spectroscopic measurements thus provide direct experimental evidence about such extension of flat bands in RG multilayers.

We then discuss the higher energy states induced by the remote-band edges. Figure 2c (lower panel) shows the energy separation between positive and negative energy states for the first (red arrows) and second (green arrows) remote-band peaks as a function of layer number. The energy separation decreases with the number of layers, indicating a smaller gap between remote bands and thus providing the enhanced screening effects from these bands in thicker RG. Theoretically, the energy gap between the remote-band states in RG is primarily determined by the number of layers and the nearest-neighbor interlayer coupling strength $\gamma_1$—a key parameter that determines the electronic properties in the tight-binding model with the Slonczewski-Weiss-McClure (SWMcC) parametrization[40,41]. Therefore, for our layer-definite RG, we can accurately extract the hopping parameter $\gamma_1$ through fitting the obtained positions of the remote-band states with the calculated bands (Fig. 2d,e). The obtained $\gamma_1$ ranges from 0.36 eV to 0.5 eV (upper panel in Fig. 2c), which is consistent well with recent transport results in trilayer and tetralayer RG[29,33], and also comparable to those reported in Bernal-stacked graphene[41]. Remarkably, $\gamma_1$ exhibits an increase with increasing thickness, indicating an enhanced interlayer interaction in more layers. Theoretically, $\gamma_1$ depends exponentially on the interlayer distance of graphene layers—the smaller interlayer distance could result in an enhanced $\gamma_1$. We did observe a decreasing trend in the interlayer spacing as the layer number increased through AFM measurements of the thickness; however, there is a relatively large margin of error in the measurements (see Supplementary Fig. 6). One possible explanation for the observed variation in the interlayer coupling could be the screening effects with a finite screening length in multilayer graphene[42,43]. According to the Thomas-Fermi theory, the screening length is estimated as $\lambda_s = 1/\sqrt{e^2\rho/\varepsilon}$ where $\rho$ is DOS. In the RG with more layers, the DOS at the CNP and in the remote bands are enhanced (see Fig. 2a,d,e). Thus, the screening length decreases in thicker RG, leading to a reduction of the interlayer distance and an increase in the interlayer interactions[44].

The observed sharp DOS peaks of the flat-band states in RG multilayers inspire the emergence of electronic correlations when these bands are partially occupied. To examine the correlation effects of flat bands, we conducted filling-varied STS measurements by combining two methods: applying a back gate voltage ($V_g$) to the device (see Fig. 1a) and detecting large RG regions with varying doping[30,45,46]. We first focus on the former case. The back gate can adjust the doping of RG, achieving in situ tuning of the filling state of the flat band. Figure 3a shows the evolution of the flat-band DOS in the 3L RG with $V_g$ measured at 77 K. The flat band exhibits a gradual transition from full filling ($V_g$ = +47 V) to slight filling ($V_g$ = -46 V), accompanied by the substantial modification of the DOS peak shape. When the flat band is occupied ($V_g$ > +20 V) or close to empty ($V_g$ < -45 V), the spectra show a single peak with weakly changed width (~50 meV for occupied states; Fig. 3b,c) as $V_g$ varies. However, the flat-band peak is dramatically broadened (up to ~90 meV; Fig. 3c) when the Fermi level lies within the flat band, and it splits into two peaks around half filling with filling-dependent spectral weights (Fig. 3a inset and Fig. 3b,d). Such remarkable filling-dependent flat-band peak broadening, splitting, and spectral weight redistribution are also observed in other RG samples with different layers (see Fig. 3e for data of the 6L RG where spectra were measured in differently doped regions at $V_g$ = 0 V, and see Supplementary Figs. 7-10 for more data). The above spectroscopic behaviors deviate from the single-particle picture (Fig. 2d,e) and directly demonstrate the emergence of interaction-induced correlated states in RG even under liquid nitrogen temperature, which closely resemble the spectral characteristics of doped flat-band peak in magic angle twisted graphene obtained at lower temperatures[47-50].

Next, we investigate the layer dependence and spectral splitting energies of the observed correlated states. Figure 4 shows the layer-dependent *dI/dV* spectra at nearly half-filling of the flat-band peak (Fig. 4a,b) and their corresponding peak-to-peak splitting energies (Fig. 4c). Note that these data were all measured in the flat regions of RG to avoid any possible extrinsic effects (see Supplementary Fig. 8 more data). The correlation-induced spectral splitting of the flat-band peak is evident in all measured 3L–9L RG. Intriguingly, the splitting energies exhibit significantly large values, ranging from ~50 meV to ~80 meV. The spectral splitting of the flat band has recently been observed in the RG system under liquid helium temperature, but with much smaller magnitudes. For example, STS study on *ABC*-stacked 4L region of twisted double-bilayer graphene measured a ~10 meV peak-to-peak splitting of the flat band[31]; and the spectroscopic measurement of multilayer RG placed on $SiO_2$ substrate resulted in a maximum flat-band splitting of ~40 meV with a critical temperature of ~20 K[30]. The RG samples studied in these two spectroscopic experiments differ from ours: the former is the moiré ABCA domain supported on $hBN/SiO_2$, and the latter lacks hBN, which may account for the deviations in the flat-band splitting. Besides, the transport experiments conducted on RG[32-34,51,52] also revealed either substantially smaller gaps or lower critical temperatures at the CNP compared to our case. This can be understood because transport measurements record average results across the global sample region, while STS detects highly local information. Additionally, it has been found that the effective transport gap of RG is smaller than anticipated due to a finite band overlap[51]; however, STS demonstrates sensitivity towards intrinsic gaps. We note that an 80 meV transport gap—comparable to our largest splitting—was previously observed up to 40 K in suspended RG devices at half filling[53], suggesting the great potential for realizing significant correlation effects in the RG system. The large correlation-induced splitting of the flat band observed at liquid nitrogen temperature here thus confirms this possibility and directly demonstrates the existence of strongly electronic interactions in multilayer RG.

The splitting energy of the flat band initially rises and then falls as the RG becomes thicker, reaching its maximum in 6L (Fig. 4c). This behavior indicates a remarkable layer dependence of the correlation effects in RG multilayers. For RG, as the number of layers increases, the flat bands host a larger DOS at the CNP (as demonstrated in Fig. 2), which promotes stronger electronic interactions due to increased electron instability. However, thicker layers also result in greater screening effects on flat-band electrons from the virtual excitations of particle-hole pairs (see Fig. 2), which will weaken the correlations. Consequently, the competition between these two factors will result in a maximum interaction strength at certain critical thickness in RG. Our spectroscopic experiment demonstrates that such critical thickness is six-layer, consistent with recent theoretical predictions[26,27]. It is worth noting that the splitting energies of the flat bands in 4L–9L RG are all enhanced compared to that in 3L, suggesting stronger correlation effects in these materials. The enhanced flat-band splitting observed in multilayer RG with $3 < N < 10$ therefore establishes a series of candidates for the investigation of strongly correlated phenomena.

To explore the nature of the observed correlated states as well as verify their layer-dependent behavior, we performed theoretical calculations within a mean-field Hartree-Fock approximation. The calculated Hartree-Fock DOS of the 6L RG are shown in Fig. 3f. A good correspondence with experiments is observed. Our theoretical calculations indicate that the correlated state emerged at the CNP (i.e., nearly half filling of the flat bands) is a layer antiferromagnetic (LAF) insulating state, and the correlated phases at slightly doped regions (e.g., at the 1/4 or 3/4 filling) flavor spin or valley polarized states (see Methods for details). To further confirm this proposal, we calculated the gap size $\Delta$ of the LAF insulating state at the CNP for different layers. The obtained results of $\Delta$ for $N < 10$ are shown in Fig. 4d. The calculated LAF gap exhibits a rise-fall variation and shows a maximum at 6L RG, which agrees well with our experimental results in terms of both the dependence on layer and numerical value. We note that previous density functional theory has also calculated nearly the same layer-dependent

LAF states in RG systems with comparable gap sizes[27].

We have examined the evolution of RG band structures and their correlated phases as the number of layers varies through STM/STS measurements. The discovered layer-dependent flat bands and interlayer hopping strength offer crucial information of the basic band structures of multilayer RG. Especially, we found layer-enhanced correlated states that persist at liquid nitrogen temperature with the maximum interaction strength in 6L. The observations of layer-dependent correlated states in $N < 10$ RG multilayers unveil several intriguing aspects, which can motivate further studies. 1) We clearly determined the layer dependence of the LAF state at the CNP, which had previously exhibited an elusive behavior in RG (3L[32,52], 4L[31,33,53], 5L[34] and thick layers[30,51]); 2) The emergence of such evident correlated states at liquid nitrogen temperature is remarkably surprising, albeit potentially influenced by our local measurements. In view of this finding, the multilayer RG represents a promising system that can host highly accessible collective phenomena against thermal fluctuation; 3) Superconductor in RG was only previously discovered in 3L with slight doping[28]. Our results demonstrate stronger correlation effects in $3 < N < 10$ RG, accompanied by the manifestation of many-body behaviors in similarly doped regions within the multilayers, thereby presenting a rich and simple material platform that holds immense potential for investigating robust or unconventional superconductivity[21].

**Methods**

**Sample fabrications.** We fabricated the RG/hBN devices using van der Waals stacking technique. Firstly, we prepared multilayer RG and hBN flakes (42–52 nm thick) on silicon substrates coated with a 285 nm oxide layer through mechanical exfoliation from bulk crystals. The *ABC* domains of RG were identified by Raman spectroscopy with a 532 nm laser. The sample thickness was determined by atomic force microscope. Then we picked up the RG flake using a handle, which is composed of a polyvinyl alcohol (PVA) film and a Polydimethylsiloxane (PDMS) stamp stacked on a glass slide. Subsequently, the RG flake was stacked onto the hBN flake on $SiO_2$/Si. The hBN flake was annealed at 300 °C for 3 h in a mixture of hydrogen and argon before assembling with RG flake. The graphene layers and hBN layers are intentionally misaligned during the fabrication process to prevent the introduction of large moiré superlattices. The PVA film was removed by dissolution in deionized water. Finally, we made electrical contact to graphene by deposition of 50–70 nm Au (or Cr/Au) electrodes through a mask. The devices were characterized again by Raman spectroscopy to confirm the preservation of *ABC* stacking before being transferred to the STM chamber.

**STM/STS measurements.** STM/STS measurements were performed in three low-temperature (liquid nitrogen) STM systems (CreaTec and Unisoku) under ultrahigh vacuum (~$10^{-11}$ Torr) and constant-current mode. Similar spectroscopic phenomena of RG were observed in all STM systems, indicating the robustness of our results. The STM topographic images were calibrated against the standard graphene lattice, Si(111)-(7×7) lattice, and Au(111) surface. The STS measurements were measured using a standard lock-in technique (AC bias modulation: 793 Hz and 10–20 mV). The electrochemically etched $Pt_{0.8}Ir_{0.2}$ and W tips were used in the measurements. The sample was located using a long-distance microscope with micron-scale resolution.

**Mean-field calculations.** The Hamiltonian of RG multilayer including noninteracting section and electron-electron Coulomb interaction is

$$H = \sum_{k,\alpha,ij} \psi_{\alpha i}^+(k) h_{ij}^\tau(k) \psi_{\alpha j}(k) + \frac{1}{2\Omega} \sum_{k,k',q} V_q \psi_{\alpha i}^+(k+q) \psi_{\beta j}^+(k'-q) \psi_{\beta j}(k') \psi_{\alpha i}(k)$$

where $h_{ij}^\tau(k)$ is the noninteracting Hamiltonian in the $\tau$ valley, and $i$ or $j$ denotes the carbon atom index. And $\alpha, \beta$ represents the spin or valley index. We adopt the dual-gate screened interaction potential $V_q = \frac{2\pi k_e}{\epsilon_r} \frac{\tanh(|q|d)}{|q|}$ in our calculations. Here $d$ is set to 50 nm and $k_e$ = 1.44 eV nm is the Coulomb constant. To investigate the screening effect, we use the constrained random-phase approximation (cRPA) in the different multilayer graphene systems. The interaction potential becomes $V_q^{cRPA} = \frac{V_q}{1+\chi V_q}$ where $\chi$ is the susceptibility obtained by cRPA calculations. We solve the total Hamiltonian including on-site Hubbard interaction ($U$ = 5 eV) by Hartree-Fock approximation with the 96×96 moment grid, and find out the ground interacting state.

The correlated insulating state that occurs at the CNP is a LAF insulating state. Because the flat bands are mainly contributed by the outermost layers, the outermost layers have the largest spin splitting due to electron-electron interactions. Each layer behaves the ferrimagnetism and thus has the non-zero net spin polarization which breaks the time-reversal symmetry. When the Fermi level is tuned away from CNP, e.g., at the 1/4 or 3/4 filling, the ground state is still dominated by correlations, flavoring a spin polarized or valley polarized state (the total energies of the correlated states are lower than those of non-interacting states, e.g., lower by 148~200 μeV per hole or per electron in the 6L RG). But under heavier electron or hole doping, the Fermi level stays far away from flat bands region, the system will transit into the non-interacting paramagnetic phase.

**Layer-dependent correlated gap.** If only local Hubbard interaction $U$ is taken into consideration, the mean-field solution about the correlated gap $\Delta$ is $1 = \frac{U\Omega}{2\pi} \int_0^\infty \frac{k dk}{\sqrt{E^2+\Delta^2/2}}$. Now we use the simplified model of low-energy bands $E = \pm(\hbar v_F k)^N / \gamma_1^{N-1}$. With this approximation, we can obtain that $\Delta^{(N-2)/N} = \frac{U\Omega}{2\pi} \int_0^\infty \frac{dx}{\sqrt{1+x^N}} (\frac{\gamma_1^{N-1}}{2\hbar^N v_F^N})^{2/N}$ (here $\Omega$ is the area of multilayer graphene unitcell). Thus, the

LAF gap $\Delta$ of $N$-layer RG is proportional to $\gamma_1^{2+\frac{2}{N-2}}$. This becomes $\gamma_1^4$ in the trilayer graphene ($N=3$), which is consistent with the previous analytical results[52]. From this formula, we can note that the stronger interlayer hopping $\gamma_1$ in the multilayer graphene can lead to the larger gap $\Delta$ contributed by the local Hubbard interaction. However, in multilayer graphene with more layers, the influence of the enhanced interlayer hopping $\gamma_1$ on the Hubbard-interaction-induced correlated gap becomes weaker.

**Enhanced screening effects.** We calculate the effective dielectric function under different $q$ in the multilayer graphene with cRPA method. In the rhombohedral trilayer graphene, the dielectric function reaches ~16 at the small $q$, which is close to the previous work[54]. While the number of multilayer graphene is increasing, the screening effect is enhancing remarkably. Especially, the dielectric function can reach the maximum value of ~30 in the 6-layer graphene (see Supplementary Fig. 12). This strong screening effect can suppress the electron-electron Coulomb interaction, and thus reduce the correlated gap of RG with more layers.

## Data availability

The data that support the findings of this study are available within the Article and its Supplementary Information. Any other relevant data are available from the corresponding author upon reasonable request.

## Code availability

The code used for the modelling in this work are available from the corresponding authors upon reasonable request.

## Acknowledgements

This work was supported by the National Natural Science Foundation of China (Grant Nos. 12174095, 12174096, 62101185, 12204164, 12304217, 11904076, 12474166,

12425405, 12404198 and 51972106), the Natural Science Foundation of Hunan Province, China (Grant No. 2021JJ20026), and the Strategic Priority Research Program of Chinese Academy of Sciences (Grant No. XDB30000000). L.J.Y. also acknowledges support from the Science and Technology Innovation Program of Hunan Province, China (Grant No. 2021RC3037) and the Natural Science Foundation of Chongqing, China (cstc2021jcyj-msxmX0381). The authors acknowledge the financial support from the Fundamental Research Funds for the Central Universities of China.

## Author contributions

Y.Z. and Y.Y.Z. fabricated the samples with the help of W.Y.L., R.J.Z. and S.M.X. Y.Z., Y.Y.Z. and L.H.T. conducted the electrode fabrications and AFM characterizations with the help of Y.T., Y.W., X.Z., X.L., Y.H. and L.L. Y.Z., Y.Y.Z. and L.J.Y. performed the STM experiments and analysed the data. T.C., Q.T., C.Z., Y.N.R. and L.H. assisted with STM measurements. S.Z. performed the mean-field calculations. H.C. performed the tight-binding calculations. W.X.W., Z.Q. and L.J.Y. supervised the experiments. L.J.Y. designed the project. Y.Z., Y.Y.Z., S.Z. and L.J.Y. wrote the manuscript with inputs from all the other authors.

## Competing interests

The authors declare no competing interests.

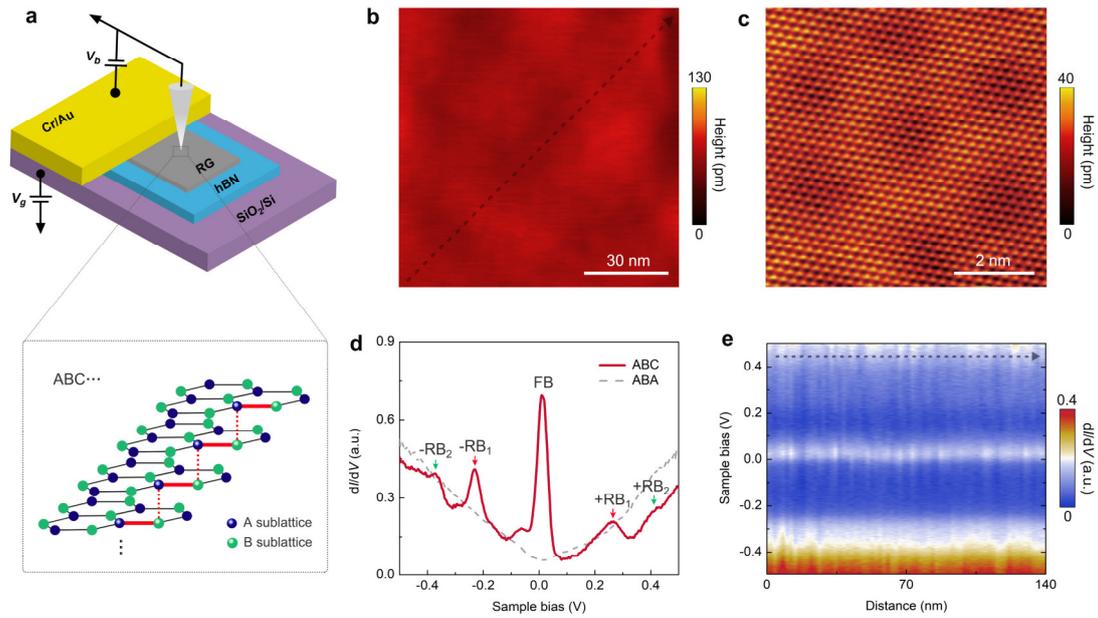

**Fig. 1 | Topography and spectroscopy of rhombohedral multilayer graphene. a,** Schematic of the STM measurement setup and rhombohedral stacking configuration. **b,** An STM topographic image (100 nm × 100 nm, $V_b$ = -0.6 V, $I$ = 0.1 nA) of a 6L RG. **c,** An atomic-resolution STM image (7 nm × 7 nm, $V_b$ = -0.6 V, $I$ = 0.1 nA) of the 6L RG showing extremely weak moiré superlattices with a small period of approximately 2.7 nm formed between the RG and hBN. **d,** Typical tunnelling conductance (*dI/dV*) spectroscopy of the rhombohedral (*ABC*) and *ABA* 6L graphene measured in the same device. FB: flat band; +RB$_1$ (-RB$_1$): the first remote conduction (valence) band; +RB$_2$ (-RB$_2$): the second remote conduction (valence) band. **e,** Contour plot of spatially resolved *dI/dV* spectra measured along the arrow in the topographic image **b**.

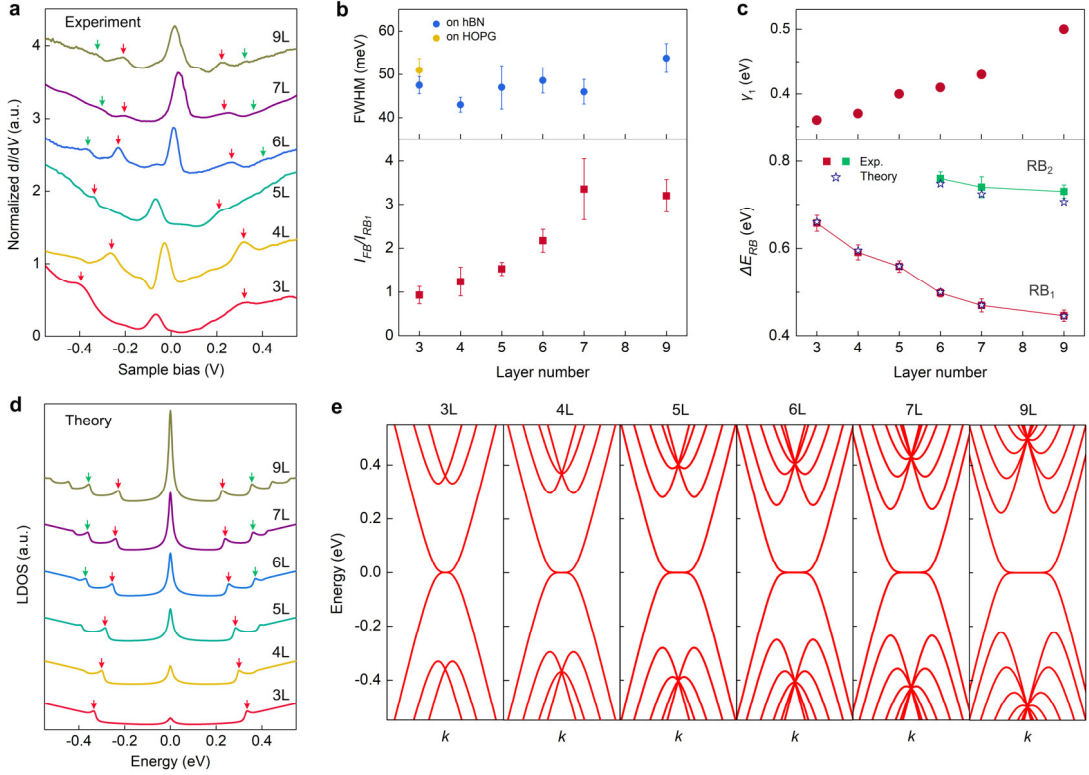

**Fig. 2 | Band structure evolution of rhombohedral multilayer graphene. a,** *dI/dV* spectra of 3L–7L and 9L RG. Arrows mark the positions of remote band peaks. Curves are vertically shifted for clarity. **b,** Upper panel: Width of the flat-band peak as a function of layer number. Blue and orange dots represent data for RG on hBN and HOPG, respectively. Lower panel: Relative intensity of the DOS peaks between the flat-band state ($I_{FB}$) and the first remote-band states ($I_{RB_1}$) in different layers. **c,** Upper panel: Obtained nearest-neighbor interlayer coupling strength $\gamma_1$ in different layers by fitting. Lower panel: Energy separation between the first (RB$_1$; red squares) and second (RB$_2$; green squares) remote bands as a function of layer number. The hollow stars are theoretically fitting results to the experimental data. Error bars in **b** and **c** represent the standard deviations (SD). Data in **b** and **c** are presented as mean values ± SD from 20 independent spectra. **d,** Calculated noninteracting LDOS on the top layer of 3L–7L and 9L RG. **e,** Single-particle band structures of 3L–7L and 9L RG obtained by the simplest tight-binding model with the nearest-neighbor intralayer coupling $\gamma_0$ = 3.16 eV. We found a negligible influence of non-nearest neighbor SWMcC hopping parameters on the determination of $\gamma_1$ (see Supplementary Figs. 4 and 5 and Supplementary Table 1).

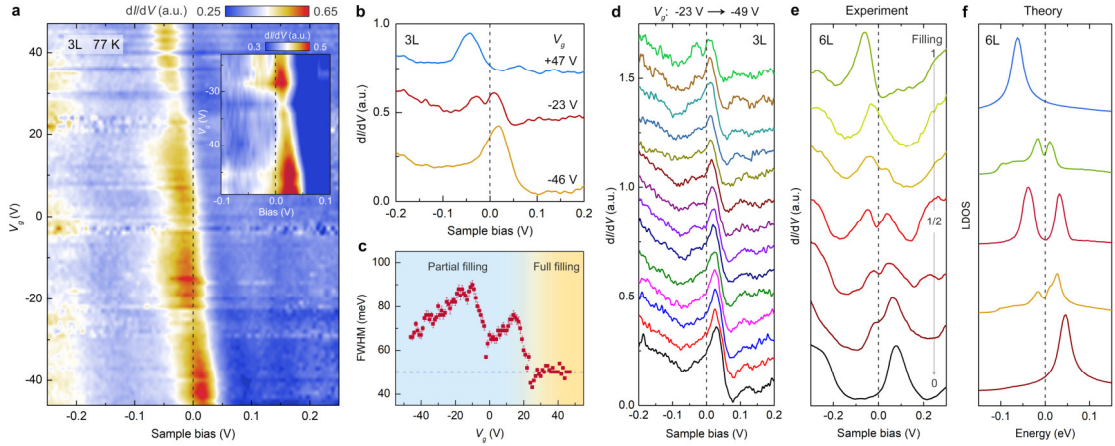

**Fig. 3 | Doping dependence of the flat-band LDOS peak. a,** $dI/dV$ spectra as a function of $V_g$ for 3L RG at 77 K. Inset, zoom-in $dI/dV$ spectra around the partial fillings of the flat band, showing splitting near half filling. **b,** Representative $dI/dV$ spectra of 3L RG for the full-filled (blue curve), nearly half-filled (red curve), and nearly empty (orange curve) flat band at different $V_g$. **c,** Width of the flat-band peak as a function of $V_g$ extracted from **a**. The width was obtained by measuring the FWHM from a single Gaussian fit to the flat-band peaks (see Supplementary Fig. 9). The blue dashed line indicates the average value of the width for full-filled flat band. Error bars represent the fitting uncertainty from the Gaussian peak fit. **d,** High-resolution $dI/dV$ spectra of the 3L RG for the partially filled flat band measured from $V_g = -23$ V to $V_g = -49$ V with a step size of 2 V. **e,** Representative tunneling spectra of 6L RG recorded at different flat regions with varying doping at $V_g = 0$ V. It shows a full variation of the flat-band filling ratio from 1 to 0 (see Supplementary Fig. 11 for extraction of the filling ratio). The flat-band peak exhibits clear splitting at partial fillings. **f,** Calculated LDOS on the top layer of 6L RG under different fillings corresponding to **e**. The curves are shifted vertically in **b** and **d-f** for clarity. The black dashed line indicates the Fermi level.

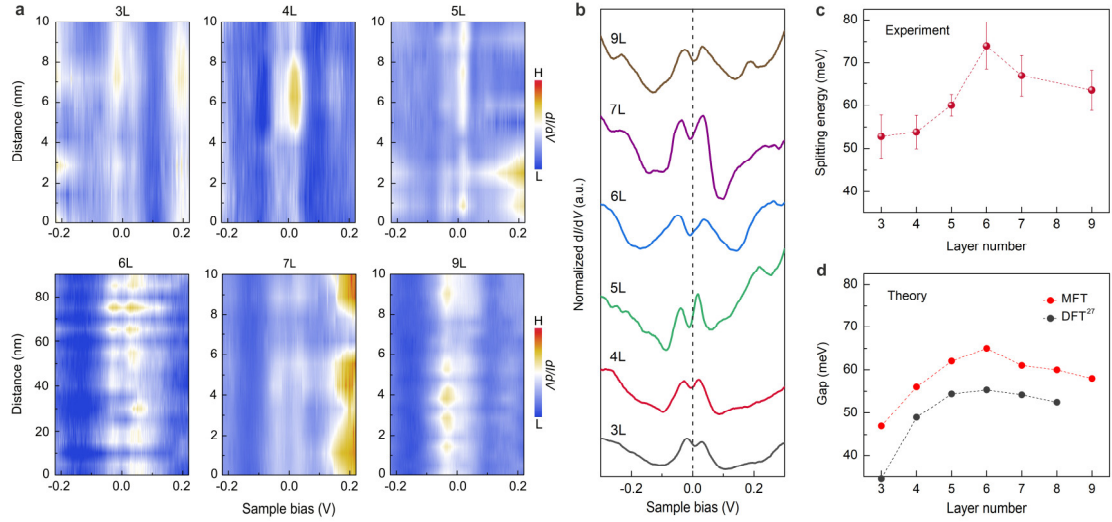

**Fig. 4 | Layer dependence of electronic correlations. a,** Spatially resolved contour plots of $dI/dV$ spectra of 3L–7L and 9L RG with split flat-band peaks recorded at $V_g = 0$ V and 77 K. The tunneling spectra were taken at flat regions of the samples. **b,** Representative STS spectra of 3L–7L and 9L RG for nearly half filling of the flat-band peak extracted from **a**. The curves are shifted vertically for clarity. **c,** Spitting energy of the flat-band peak (half filling) as a function of layer number. Data are presented as mean values ± SD from 20 independent spectra. **d,** Theoretical energy gap of the correlated state at half filling under 77 K. The red circles are our mean-field theory (MFT) results. The black circles indicate density functional theory (DFT) results extracted from literature[27].